\documentclass[prl,showpacs]{revtex4}

\usepackage{amsmath}
\usepackage{amssymb}

\usepackage{color}
\usepackage{graphicx}
\usepackage[english]{babel}

\begin{document}

\title{Influence of s-d scattering on the electron
density of states in ferromagnet/superconductor bilayer}

\author{D. Gusakova$^{1}$, A. Vedyayev$^{1}$ , O. Kotelnikova$^{1}$, N. Ryzhanova$^{1}$, A. Buzdin$^{2}$}
\address{
$^1$ Faculty of Physics, M.V.Lomonosov Moscow State University,
Moscow 119992,Russia
\\$^2$ Centre de Physique Th\'{e}orique et de Mod\'{e}lisation,
ERS-CNRS 2120, Universit\'{e} Bordeaux I, 33405 Talence Cedex,
France}

\begin{abstract}
  We study the dependence of the electronic density of states (DOS) on the
distance from the boundary for a ferromagnet/superconductor
bilayer. We calculate the electron density of states in such
structure taking into account the two-band model of the
ferromagnet (FM) with conducting s and localized d electrons and a
simple s-wave superconductor (SC). It is demonstrated that due to
the electron s-d scattering in the ferromagnetic layer in the
third order of s-d scattering parameter the oscillation of the
density of states has larger period and more drastic decrease in
comparison with the oscillation period for the electron density of
states in the zero order.
\end{abstract}


\maketitle

As it well known superconductivity creates a gap in the electronic
density of states (DOS) near the Fermi energy $E_F$. In the case
of ferromagnet/superconductor multistructures at the SC/FM
boundary the Cooper pair wave function extends from the SC into
the FM layer. The internal exchange field in a ferromagnet results
in a strong suppression of the superconducting order parameter.
Damped oscillatory dependence of the Cooper pair wave function in
FM hints that a similar damped oscillatory behavior may be
expected for the variation of the DOS due to proximity effect. For
SC/FM bilayer, for example, this question has been studied in
\cite{Buzdin2000}. It was shown that the DOS oscillations in FM
near the boundary are described by the expression

\begin{equation}
N_f(\varepsilon=0)\approx N(0)\left(
1-\frac{1}{2}exp(-\frac{2x}{\xi_f})cos(\frac{2x}{\xi_f}) \right),
\end{equation}
where $\xi_f$ is the characteristic length of the superconducting
correlations decay in FM layer, $x$ is the distance from the
boundary.

In most theoretical papers \cite{Buzdin2000,Fazio1999} the
nonmonotonic behavior of the superconducting order parameter was
studied in the so-called dirty limit and for low energies, that is
using the quasiclassical Usadel equations in the context of
one-band model of ferromagnetic metal. In ref.\cite{Bergeret2002}
the more general Eilenberger equations for a larger energy range
are discussed. However, due to some simplifications, in both cases
there are evident limitations which could be avoided by solving
the system of full Gor'kov equations for the normal and anomalous
Green functions. Moreover, when calculating the DOS in 3d
ferromagnetic metal one should take into account the s-d electron
scattering which is the main scattering mechanism responsible for
the kinetic properties of 3d metal due to the large DOS of d
electrons at the Fermi surface.

As it was substantiated above we study the influence of the s-d
scattering of the conducting electrons on the DOS in the
ferromagnetic layer taking into account the two-band model of the
FM in contact with the SC using Green function approach. The FM is
assumed to be a 3d ferromagnetic metal with two types of electrons
- conducting s electrons and almost localized d electrons. The SC
is a simple s-wave superconductor.

 We consider the planar semi-infinite
geometry with an FM to the left of z=0 and an SC to the right. The
Hamiltonian in the F layer is
\begin{equation}
\begin{array}{c}
  \widehat{H}=\sum_{\alpha}\int d\textbf{r}\psi_\alpha^{+s}\xi_s
\psi_\alpha^s +\sum_{\alpha}\int d\textbf{r}\psi_\alpha^{+d}\xi_d
\psi_\alpha^d+\\
   +\sum_\alpha\sum_n\int
d\textbf{r}\delta(\rho-\rho_n)\gamma_n[\psi_\alpha^{+s}\psi_\alpha^d
 +\psi_\alpha^{+d}\psi_\alpha^s
], \\
\end{array}
\end{equation}
in the S-layer is
\begin{equation}
\widehat{H}=\sum_{\alpha}\int d\textbf{r}\psi_\alpha^{+s}\xi_s
\psi_\alpha^s+\int
d\textbf{r}[\Delta(\textbf{r})\psi_\uparrow^{+s}\psi_\downarrow^{s+}+c.c.],
\end{equation}
where $\psi_\sigma^{+s(d)}$ and $\psi_\sigma^{s(d)}$  are the
creation and annihilation
 operators for s(d) electrons with spin $\sigma$, $\Delta(\textbf{r})$
  is the superconducting order parameter,
  $\xi_{s(d)}$ are the dispersion
functions for the s(d) electrons, $\gamma_n$ is the electron
random s-d scattering parameter. In the calculations we use the
averaged over impurity configurations value of
$\gamma=\overline{\gamma_n}$ and
$(\Delta\gamma)^2=\overline{\gamma_n^2}-\overline{\gamma_n}^2$.
The normal and anomalous Green functions of the one-band model are
$G^{ss}(x_1,x_2)=-\langle T_\tau \psi^s(x_1)\psi^{+s}(x_2)\rangle$
and $F^{ss}(x_1,x_2)=-\langle T_\tau
\psi^{+s}(x_1)\psi^{+s}(x_2)\rangle$, where $x=(\tau,\textbf{r})$
is a four-component vector and the creation and annihilation
operators are associated with s conducting electrons. The two-band
model requires us to take into account similar Green functions of
d electrons and the Green functions responsible for the s-d
electron scattering. Then in the FM layer with s-d scattering the
full Green function can be written as the following matrix in the
s-d space:
\begin{equation}
\widehat{G}=\left(%
\begin{array}{cccc}
  \widehat{G}^{ss} & \widehat{G}^{sd} & -\widehat{F}^{+ss} & -\widehat{F}^{+sd} \\
  \widehat{G}^{ds} & \widehat{G}^{dd} & -\widehat{F}^{+ds} & -\widehat{F}^{+dd} \\
  -\widehat{F}^{ss} & -\widehat{F}^{sd} & \widehat{G}^{+ss} & \widehat{G}^{+sd} \\
  -\widehat{F}^{ds} & -\widehat{F}^{dd} & \widehat{G}^{+ds} & \widehat{G}^{+dd} \\
\end{array}%
\right),
\end{equation}
in the SC layer the same matrix has the form:
\begin{equation}
\widehat{G}=\left(%
\begin{array}{cccc}
  \widehat{G}^{ss} & 0 & -\widehat{F}^{+ss} & 0 \\
 0 & 0 & 0 &0 \\
  -\widehat{F}^{ss} & 0 & \widehat{G}^{+ss} & 0 \\
0 & 0 & 0 &0 \\
\end{array}%
\right),
\end{equation}
In the spin space, each of the functions $\widehat{G}^{ij}$  and
$\widehat{F}^{ij}$ is also the two by two matrix:
\begin{equation}
\widehat{G}^{ss}=\left(%
\begin{array}{cc}
 G_{\uparrow\uparrow}^{ss} & 0 \\
  0 & G_{\downarrow\downarrow}^{ss} \\
\end{array}%
\right), \widehat{F}^{ss}=\left(%
\begin{array}{cc}
  0 & F_{\uparrow\downarrow}^{ss} \\
  F_{\downarrow\uparrow}^{ss} & 0 \\
\end{array}%
\right).
\end{equation}
With the account of the term responsible for the s-d electron
scattering in (1) the system of the Gor'kov equations for normal
and anomalous Green functions become rather complicated. Such a
system is no longer linear and requires the perturbation solution
over the parameter $\gamma$ which is small in comparison with the
electron kinetic energy.

In the zero order on $\gamma$ we have the anomalous function F
only of the s electrons due to the boundary conditions with the
S-layer. In order to obtain all other functions in (3) the next
approximations have to be investigated. In the second order on the
s-d electron scattering parameter the Gor'kov equations contain
$F^{ds}$ function which is connected with $F^{ss}$ by means of
$\gamma$. Anomalous $F^{sd}$ and $F^{dd}$ functions in the F-layer
originate from their connection with $G^{dd}$ in higher order on
$\gamma$. We are interested in the third order on the s-d
scattering parameter $\gamma$ as in this case the solution of the
Gor'kov equations has the form of plane waves with the exponents
responsible for the change of the electron density of states
oscillation period due to s-d scattering.

We consider the full Green function as the sum of the components
of the corresponding order on $\gamma$:
$G=G_{(0)}+G_{(1)}+G_{(2)}+\ldots$ and
$F=F_{(0)}+F_{(1)}+F_{(2)}+\ldots$. By carrying out the Fourier
transformation in the plane perpendicular to the $z$ axes (using
the quasi momentum representation in the $xy$ plane) in the third
order on $\gamma$ we get the following system of Gor'kov equations
in the F-layer:
\begin{equation}
\left\{
\begin{array}{c}
  \widehat{\Xi}_s^\uparrow \cdot G_{\uparrow\uparrow}^{(3)ss}(z,z',\omega)
  -\gamma\cdot G_{\uparrow\uparrow}^{(3)ds}(z,z',\omega)=0, \\ \\
  -\gamma \cdot G_{\uparrow\uparrow}^{(3)ss}(z,z',\omega)
  +\widehat{\Xi}_d^\uparrow \cdot G_{\uparrow\uparrow}^{(3)ds}(z,z',\omega)=\Upsilon_F,
  \\ \\
  -\widehat{\Xi}_s^\uparrow\cdot F_{\downarrow\uparrow}^{(3)ss}(z,z',\omega)+
  \gamma\cdot F_{\downarrow\uparrow}^{(3)ds}(z,z',\omega)=0, \\ \\
  -\gamma\cdot F_{\downarrow\uparrow}^{(3)ss}(z,z',\omega)+
  \widehat{\Xi}_d^\downarrow\cdot F_{\downarrow\uparrow}^{(3)ds}(z,z',\omega)=\Upsilon_G.
  \\
\end{array}
\right.
\end{equation}
Here\begin{equation}
\begin{array}{r}
  \widehat{\Xi}_s^{\uparrow(\downarrow)}= i\omega\pm
\frac{1}{2m_s}\left(\frac{\partial^2}{\partial
z^2}-\chi^2\right)\pm\frac{k_{Fs}^{\uparrow(\downarrow)2}}{2m_s}\mp\\\mp
x_0 (\Delta \gamma)^2
G_{\uparrow\uparrow(\downarrow\downarrow)}^{(0)dd}(z,z), \\ \\
  \widehat{\Xi}_d^{\uparrow(\downarrow)}= i\omega\pm
\frac{1}{2m_s}\left(\frac{\partial^2}{\partial
z^2}-\chi^2\right)\pm\frac{k_{Fs}^{\uparrow(\downarrow)2}}{2m_s},\\
\\
  \Upsilon_F= -\left[ (\Delta\gamma)^2 F_{\uparrow\downarrow}^{(0)+ss}(z,z)\right]
   \cdot F_{\downarrow\uparrow}^{(2)ds}(z,z',\omega),\\ \\
  \Upsilon_G= -[ (\Delta\gamma)^2 F_{\downarrow\uparrow}^{(0)ss}(z,z)]
  \cdot G_{\uparrow\uparrow}^{(2)ds}(z,z',\omega),\\
\end{array}
\end{equation}
where $k_{Fs(d)}^{\uparrow(\downarrow)}$ is the Fermi impulse of
s(d) electrons with the spin up (down); $k_{Fsup}$  the electron
Fermi impulse in the S-layer; $m_{s(d)}$ the mass of s(d)
electrons; $\omega$ the energy parameter; $\varepsilon_F$ the
Fermi energy; $x_0$ the impurity concentration. The functions
$G^{(0)}$,$G^{(2)}$,$F^{(0)}$, and $F^{(2)}$ are considered to be
known from the previous order on $\gamma$. In particular the
normal diagonal $(z=z')$ Green function of the zero order
$G^{(0)}$ has the form
\begin{equation}
G^{(0)ss}(z,z)=const_1+const_2 e^{2izk_{(0)1}},\text{where}
\end{equation}
the coefficients

 $const_{1,2}$$=$$const_{1,2}(k_{(0)1},k_{(0)2},k_3,\omega,\Delta,(\Delta\gamma)^2)$\\
are the functions of
\begin{equation}
k_{(0)1}=\sqrt{k_{Fs}^{\uparrow
2}-\chi^2+2im_s\omega-2m_sx_0(\Delta\gamma)^2G_{\uparrow\uparrow}^{dd}(z,z)},
\end{equation}
\begin{equation}
k_{(0)2}^*=\sqrt{k_{Fs}^{\downarrow
2}-\chi^2-2im_s\omega+2m_sx_0(\Delta\gamma)^2G_{\downarrow\downarrow}^{+dd}(z,z)},
\end{equation}
superconducting order parameter $\Delta$, energy parameter
$\omega$, and s-d electron scattering parameter $\gamma$. One can
see that in the zero order on $\gamma$ the electron density of
states of s electrons near the boundary with the S-layer has a
rapid oscillation with the period proportional to
$\pi/k_{Fs}^\uparrow$ and exponential decaying to the bulk value
at large $z$. In this case, the conducting electrons undergo the
ordinary and Andreev reflection at the border F/S which leads to
this rapid oscillations. Another type of oscillations due to s-d
electron scattering appears in higher order on $\gamma$ when the
relationship between the normal s electron Green function  and
anomalous takes an explicit form. Then we get the superposition of
the ordinary rapid oscillations due to Andreev reflection on the
interface and slower ones with the period proportional to the
reversed difference of the Fermi momenta of s electrons with spin
up and spin down.

In the S-layer the system of Gorkov equations for the normal and
anomalous Green functions keeps its usual form:
%
\begin{equation}
\left\{\begin{array}{l}
     \left[i\omega+\frac{1}{2m_s}\left(\frac{\partial^2}{\partial z^2} -\chi^2\right)+\frac{k_{Fsup}^{2}}{2m_s} \right]
   \times\\\times G_{\uparrow\uparrow}^{ss}(z,z',\omega)+
                            +\Delta\cdot F_{\downarrow\uparrow}^{ss}(z,z',\omega)=\delta(z-z'),
   \\ \\
  -\Delta^*\cdot G_{\uparrow\uparrow}^{ss}(z,z',\omega)-\\ -\left[i\omega-\frac{1}{2m_s}\left(\frac{\partial^2}{\partial z^2}
  -\chi^2\right)-\frac{k_{Fsup}^{2}}{2m_s} \right]\cdot F_{\downarrow\uparrow}^{ss}(z,z',\omega)=0.\\
\end{array}
\right.
\end{equation}
The Green functions  $G^{ds}$ and $F^{ds}$  in S-layers are
identically zeros.

In the third order of $\gamma$ the diagonal function
$G_{(3)}^{ss}(z,z,\chi)$ which serves to calculate the DOS in the
FM layer represents the following sum of the plane waves with the
different wave vectors:
\begin{equation}
\begin{array}{r}
  G_{(3)}^{ss}(z,z,\chi)=b_1e^{-2iz(k_4^*-k_1)}+b_2e^{-iz(k_4^*+k_5^*-2k_1)} \\
 +b_3e^{-iz(k_4^*-2k_2-k_1)}+b_4e^{-iz(k_4^*-k_1-2k_2)}, \\
\end{array}
\end{equation}
where the coefficients are the functions of the $b_{i} = $  $
\gamma(\Delta\gamma)^2
f_i(k_1,k_2,k_3,k_3^*,k_4^*,k_5^*,\omega,\Delta)$, $\omega$ is the
energy parameter. As one can see the first term in (7) contains
the exponent with the argument proportional to the doubled
difference of the Fermi impulses of s electrons with the opposite
spin directions $(k_{Fs}^\uparrow-k_{Fs}^\downarrow)$. That is,
this term gives the increase of the characteristic period of
oscillations due to s-d scattering. The effective exponential
decaying is defined by the exponent
$e^{-z(\frac{1}{l^\uparrow}+\frac{1}{l^\downarrow})}$, where
$l^{\uparrow(\downarrow)}=k_{Fs}^{\uparrow(\downarrow)}/(2m_sx_0(\Delta\gamma)^2
ImG_{\uparrow\uparrow(\downarrow\downarrow)}^{dd})$ is the free
path of the s electron with the spin up(down) in ferromagnetic
layer, $m_s$ is the mass of the s electron, $x_0$ is the impurity
density, $G_{\uparrow\uparrow(\downarrow\downarrow)}^{dd}$ is the
Green function of d electrons. In order to calculate the DOS we
use the imaginary part of the diagonal Green function: $
\rho^s(z)=\tiny{\frac{1}{2\pi}}Im\int\chi d\chi G^{ss}(z,z,\chi)$.
 The graphical dependencies of the DOS of s electron  on the
distance from the boundary in the ferromagnetic layer in the zero
order and in the third order for the different values of $\gamma$
and fixed $(\Delta\gamma)^2$ are presented in Fig. 1(a) and 1(b),
correspondingly. The oscillations of the third order term in the s
electron density of states have larger period and more drastic
decrease in comparison with the oscillation period for the
electron density of states in the zero order. This fact evidences
the influence of the s-d electron scattering of conducting
electrons on their density of states in the ferromagnetic layer.

\begin{figure}[h]   
  \includegraphics[width=0.5\textwidth]{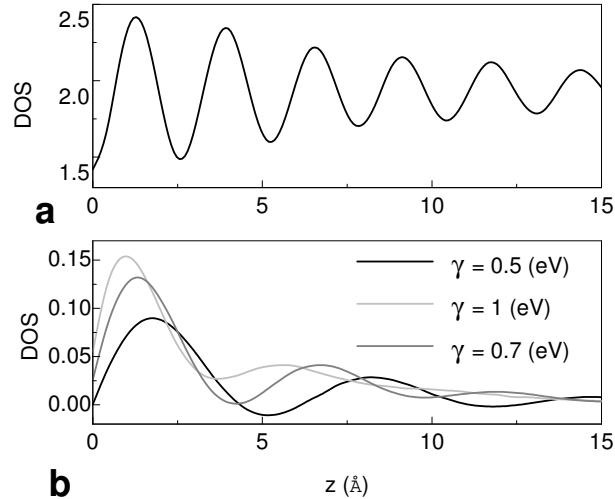}\\
  \caption{(a) DOS as a function of the distance $z$ from the
  boundary in FM in the zero approximation. (b) The additional
 term to the DOS as a
  function of the distance $z$ from the
  boundary in FM for $\omega=0$ and three different
values of s-d electron scattering parameter $\gamma$
($k_{Fs}^{\uparrow}=1.2$ $\AA^{-1}$, $k_{Fs}^{\downarrow}=0.42$
$\AA^{-1}$, $(\Delta\gamma)^2=0.1$ $eV^2 $).}\label{fig:1}
\end{figure}

\begin{figure}[h]
  \includegraphics[width=0.4\textwidth]{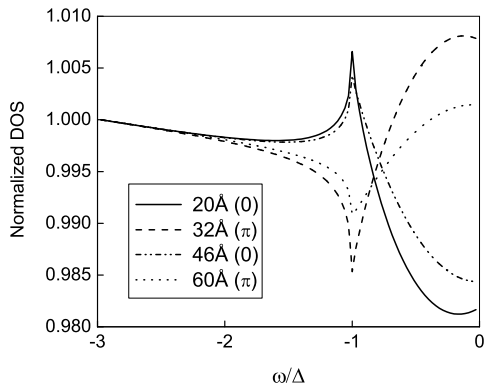}\\
  \caption{Energy variation of the DOS in FM for different distances $z$
 from the boundary. Zero energy corresponds to the Fermi level,
   energy gap $\Delta=1.4$ meV, exchange field $h=10$ meV.}\label{fig:2}
\end{figure}

On the FM side, the proximity effect induces the superconducting
order parameter. Increasing the distance from the interface it
displays a damped oscillation and changes sign from positive to
negative. Usually the states corresponding to a positive sign of
superconducting order parameter are called the "0 state" and those
corresponding to a negative sign of order parameter the "$\pi$
state". In ref.\cite{Guoya2002} Guoya \textit{et al.} develop a
quantum-statistical approach based on the McMillan and BTK
theories to calculate the DOS dependence on the energy. They show
that the DOS in FM displays a maximum at the energy-gap edge and a
minimum at the Fermi level exhibiting the SC-like shape for the
"0" state, and flipped shape for the "$\pi$" state.

Figure 2 shows the result of our calculations of the DOS in FM
using the Gor'kov equations. As it may be seen the DOS exhibits
the same as in ref. \cite{Guoya2002} peak-dip behavior near the SC
gap for "0" and "$\pi$" states. With increasing the distance from
the interface, the SC-like behavior of the DOS in F-layer
gradually disappears. Such two different DOS shapes in FM near the
FM/SC boundary for the "0" and "$\pi$" states have been observed
experimentally by Kontos \textit{et al.} \cite{Kontos2000} in the
measurements of the DOS by planar-tunnelling spectroscopy in
$Al/Al_2O_3/PdNi/Nb$ junctions.

On the SC side the superconducting order parameter is diminished
near the FM/SC interface (see Fig. 3) and owing to the proximity
effect the superconductivity in the SC near the interface also
becomes gapless. Far from the boundary the DOS in SC takes it
usual shape as in bulk superconductor.

\begin{figure}[h]
  \includegraphics[width=0.4\textwidth]{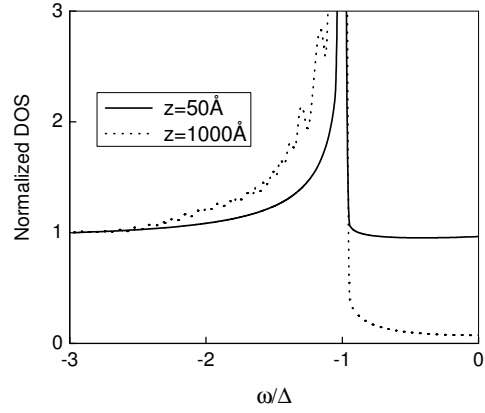}\\
  \caption{Energy variation of the DOS in the SC near (solid line)
   and far (dotted line) from the boundary.}\label{fig:3}
\end{figure}


This work was supported by the Russian Foundation for Basic
Research (grant N 04-02-16688 a).

\end{document}